\documentclass[aps,prb,showpacs,notitlepage,floatfix,twocolumn,superscriptaddress]{revtex4-2}
\usepackage{amssymb}
\usepackage{amsmath}
\usepackage{amsthm}
\usepackage{amsfonts}
\usepackage{graphicx}
\usepackage{braket}
\usepackage{mathtools}

\usepackage[dvipsnames]{xcolor}
\usepackage[titletoc]{appendix}

\newcommand\myshade{85}
\colorlet{myurlcolor}{TealBlue}

\usepackage[colorlinks,linkcolor=myurlcolor!\myshade!black,anchorcolor=myurlcolor!\myshade!black,citecolor=myurlcolor!\myshade!black,urlcolor=Magenta]{hyperref}

\graphicspath{{Figures/}}

\def\bea{\begin{equation}\begin{aligned}}
\def\eea{\end{aligned}\end{equation}}
\newcommand{\be}{\begin{equation}}
\newcommand{\ee}{\end{equation}}

\begin{document}

\title{Entanglement patterns of quantum chaotic Hamiltonians with a scalar U(1) charge}

\author{Christopher M. Langlett}
\affiliation{Department of Physics \& Astronomy, Texas A\&M University, College Station, TX 77843, USA}

\author{Joaquin F. Rodriguez-Nieva}
\affiliation{Department of Physics \& Astronomy, Texas A\&M University, College Station, TX 77843, USA}

\begin{abstract}
  Our current understanding of quantum chaos in many-body quantum systems hinges on the random matrix theory~(RMT) behavior of eigenstates and their energy level statistics. Although RMT has been remarkably successful in describing `coarse' features of many-body quantum Hamiltonians in chaotic regimes, such as the Wigner-Dyson level spacing statistics or the volume-law behavior of eigenstate entanglement entropy, it remains a challenge to describe their `finer' features, particularly those arising from spatial locality. Here, we show that we can accurately describe the statistical behavior of {\it eigenstate ensembles} in many-body Hamiltonians by using pure random states with physical constraints that capture the essential features of the Hamiltonian, specifically spatial locality and symmetries. We demonstrate our approach on local spin Hamiltonians with a scalar U(1) charge. By constructing ensembles of constrained random states that account for two {commuting} scalar charges playing the role of energy and magnetization, we describe the patterns of entanglement of mid-spectrum eigenstates {\it beyond} their average volume-law behavior, including $O(1)$ corrections and fluctuations, analytically and numerically. When defining the correspondence between quantum chaotic eigenstates in many-body Hamiltonians and RMT ensembles, our work highlights the important role played by spatial locality in describing universal features beyond the volume-law behavior.
\end{abstract}
\maketitle

\noindent{\bf Introduction.---}Understanding how statistical mechanics emerges in isolated many-body quantum systems has been a long-standing challenge~\cite{deutsch1991quantum,srednicki1994chaos,rigol2008thermalization,nandkishore2015many}. {The fundamental difficulty in addressing this challenge is describing how chaos emerges from unitary evolution in quantum systems. Valuable insights into this foundational question} come from random matrix theory~(RMT), which has been remarkably successful in describing `coarse' features of many-body quantum states in chaotic regimes. In particular, the widely-accepted expectation is that the eigenstates and eigenvalues of generic quantum many-body Hamiltonians (i.e., away from fine-tuned integrable limits) exhibit universal statistics described by RMT ensembles, such as the Gaussian Orthogonal Ensemble (GOE) in the context of time-reversal invariant Hamiltonians. This expectation applies to eigenspectrum properties, such as the level spacing statistics~\cite{Atas2013,atas2013joint,oganesyan2007localization} or the spectral form factor~\cite{bertini2018exact,chan2018spectral,FriedmanSpectral2019}, and to eigenstate properties, such as the thermal-like behavior of local observables~\cite{srednicki1999approach,d2016quantum,2018RPP_Deutsch,dymarsky_subsystem_2018} and the volume-law behavior of the entanglement entropy~(EE)~\cite{Garrisondoes2018,Lurenyi2019,entanglementvidmar2017,structuremurthy2019,volumebianchi2022}.

\begin{figure}[t]
    \includegraphics[width=\columnwidth]{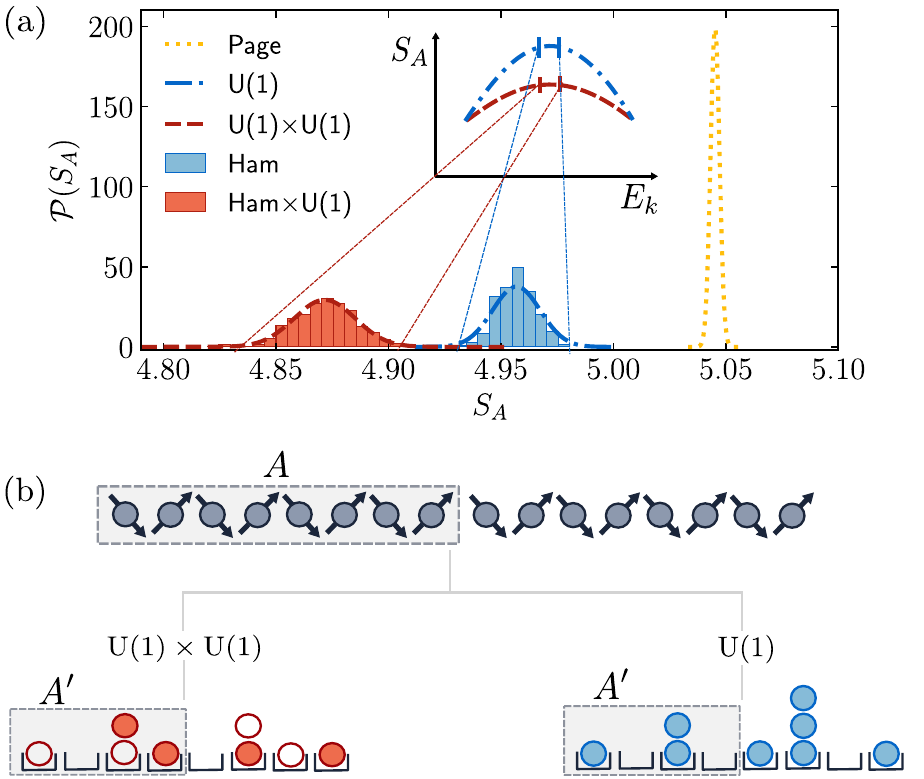}
    \caption{(a) Histograms of half-system entanglement entropy (EE) of mid-spectrum eigenstates for Hamiltonian systems with a scalar U(1) charge~(left, red) and for a Hamiltonian system without additional symmetries~(right, blue). The lines indicate the EE distribution of pure random states without constraints (dotted), random states with a U(1) constraint (dashed-dotted), and random states with two U(1) constraints (dashed).
    (b) Schematics showing the {\it coarse-graining} procedure used to generate quantum state ensembles that capture the entanglement patterns of a spin chain Hamiltonian of spin-$1/2$ qubits, shown for $L=16$. In particular, we consider a chain of $L' = L/\lambda$ qudits with local Hilbert space dimension $d' = d^\lambda$, with $\lambda=2$, and consider random states within a fixed symmetry sector for each U(1) charge. 
    Parameters used in the histograms: $L=16$, $J=1$, $\Delta=J/2$, $J_{2}=0.6$, $h_{x}=0$~[Ham$\times$U(1)], and $h_{x}=0.75$, $J_2=0$~(Ham), see Eqs.(\ref{eq:ModelHam}) and (\ref{eq:ModelHam2}).
    }
    \label{fig:Fig1}
\end{figure}

While RMT ensembles successfully describe `coarse' features of quantum states in chaotic regimes, describing their `finer' features remains a challenge. In particular, eigenstates of many-body Hamiltonians encode physical correlations that are not described by eigenstates of dense, random matrices such as GOE. In recent years, many works have aimed to capture structure in many-body quantum states that goes beyond RMT, particularly that arising from spatial locality. For example, Refs.~\cite{Garrattlocal2021,dymarskybound2022,Eigenstatewang2022,richtereigenstate2020,brenesout2021,eigenstatefoini2019,eigenstatechan2019} showed that matrix elements of local operators evaluated in the eigenbasis of the Hamiltonian remain correlated up to a particular energy scale related to the so-called Thouless time~\cite{bertini2018exact,gharibyan2018onset,chan2018solution,chan2018spectral,FriedmanSpectral2019}. 
Other works have identified systematic deviations between the EE of mid-spectrum energy eigenstates and pure random states in spatially local systems~\cite{entanglementhaque2022,huang2019universal,huang2021universal,huang2022deviation,entanglementvidmar2017,Kliczkowskiaverage2023}. In fact, spatial locality is {\it strongly} imprinted in the structure of individual eigenstates, as a single eigenstate is sufficient to reconstruct the entire Hamiltonian~\cite{qi2019determining, Garrisondoes2018}. 

Despite these works, statistical descriptions of many-body eigenstates in chaotic regimes that incorporate spatial locality as the fundamental building block still remain to be developed. The perspective that we take in this letter is to construct ensembles of pure random states that capture increasingly finer features of typical quantum states of physical Hamiltonians by imprinting symmetries on top of spatial locality. A first step in this direction was taken in our recent work \cite{rodriguez2023quantifying}, where we showed that the entanglement entropy (EE) statistics of mid-spectrum eigenstates in local many-body Hamiltonians is not described by the GOE ensemble but, instead, by a constrained ensemble that incorporates a U(1) scalar charge---energy in local systems plays the role of the scalar charge~\cite{rodriguez2023quantifying}. Similar arguments were used by Huang~\cite{huang2019universal,huang2021universal} to justify deviations from the maximal EE of eigenstates in local Hamiltonian systems. {Using this mapping between energy eigenstates and ensembles constrained by U(1) scalar charge, it is possible to derive analytic expressions for the asymptotic behavior of the EE entropy distributions, including $O(1)$ corrections and fluctuations, as calculations can be more easily done in the U(1) case.}

\begin{figure}[t]
    \includegraphics[width=\columnwidth]{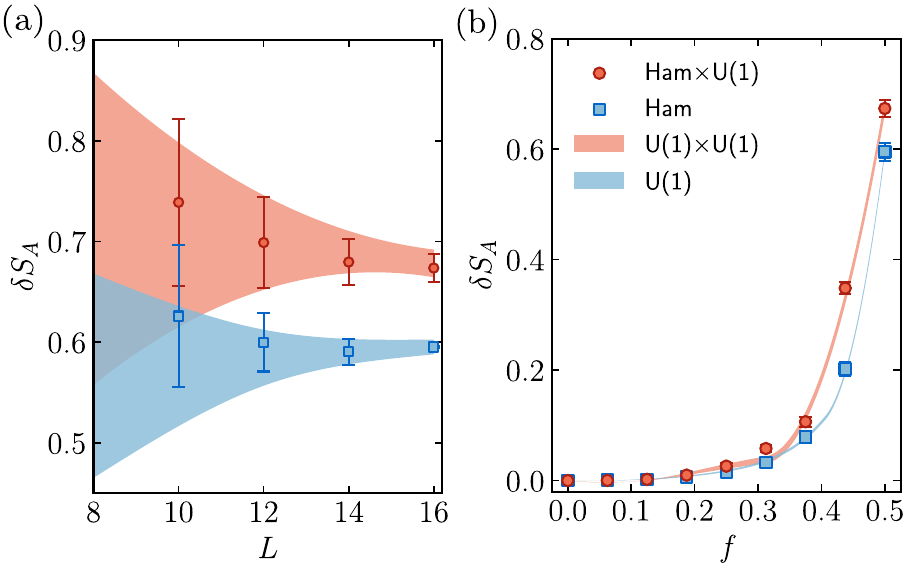}
    \caption{
    (a) Finite-size scaling of the EE distribution as a function of $L$ for fixed $f=1/2$. The data is shown relative to the maximum entropy, $\delta S_A = Lf\log(2) - S_A$. The red points indicate the average EE, and the bars indicate the standard deviation of EE. (b) Distribution of EE for mid-spectrum eigenstates as a function of $f$ for fixed $L=16$. The shaded areas indicate the regions limited by $S_A = \mu_M \pm \sigma_M$ (blue) and $S_A = \mu_{M,N} \pm\sigma_{M,N}$ (red) for the ensembles with one and two scalar constraints, respectively,  in coarse-grained systems of size $L'=[4,6,8]$ [in panel (b), the shaded regions are narrow]. Parameters used in panels (a) and (b): (red) $J=1$, $\Delta=J/2$, $J_{2}=0.6$, $h_{x}=0$ and (blue) $J_{2}=0$, $h_{x}=0.75$. For small values of $f$, the standard deviation bars are hidden behind the data points.
    }
    \label{fig:Fig2}
\end{figure}

Here we show that we can accurately describe the entanglement patterns of many-body eigenstates in local systems---including $O(1)$ corrections and fluctuations---{in more general settings. We focus specifically on the simplest case of} Hamiltonian systems with an additional U(1) scalar charge that commutes with energy, see Fig.~\ref{fig:Fig1}. Systems with a U(1) scalar charge arise in many models of experimental interest, such as magnetic systems, and systems with particle number conservation, such as interacting Fermi and Bose gases~\cite{trotzky2012probing, Tang2018thermal}. By constructing random state ensembles that account for the conservation of energy and magnetization on top of spatial locality, we accurately describe the ensemble properties of eigenstates {\it beyond} their average volume-law behavior, both analytically and numerically (Figs.~\ref{fig:Fig1} and \ref{fig:Fig2}). Importantly, our results suggest the existence of universality in the $O(1)$ corrections and fluctuations of the EE distribution which is governed by the number of conserved charges, as discussed below. 

\noindent{\bf Overview of the approach.}---Our goal is to describe the statistical properties of midspectrum eigenstates in local systems through the lens of EE,
\be
S_{A} = -\text{Tr}[\rho_{A}\log(\rho_{A})], \quad \rho_A = {\rm Tr}_B[|\Psi\rangle\langle\Psi|].
\label{eq:EE}
\ee
In Eq.(\ref{eq:EE}), $\rho_A$ is the reduced density matrix of the state $|\Psi\rangle$ when the system is bipartitioned into two subsystems of length $L_A = fL$ and $L_B=(1-f)L$, each of which has Hilbert space dimension $d_A=d^{L_A}$ and $d_B = d^{L_B}$ ($d$ being the local Hilbert space dimension). Without loss of generality, we assume $f \le 1/2$. 
The states $|\Psi\rangle$ are drawn from a small microcanonical window of size $\epsilon \ll 1 $ near the middle of the spectrum for {\it each} conserved quantity ${Q_\alpha} = \sum_{i=1}^L {q}_{\alpha,i}$, which we assume to be local (i.e., $q_{\alpha,i}$ acts on a few neighboring degrees of freedom) and commuting with each other. 

Since we are specifically interested in the {\it asymptotic} behavior of the subsystem EE (i.e., $d_A,d_B \gg 1$), the essence of our approach is based on the observation that the distribution of EE remains largely unaffected by the specific details of the local charges $Q_\alpha$ as the system scales up in size. This holds as long as the oprators ${q}_{\alpha,i}$ remain local, acting only on a few degrees of freedom. As such, we can adopt the simplest possible form for the local charges $q_{\alpha,i}$---specifically, single-site operators---that preserve all Hamiltonian symmetries when computing the asymptotic behavior of EE, including $O(1)$ corrections and fluctuations. For the remaining of the letter, we first describe the main analytical results as the number of conservation laws increases, followed by numerical results on Hamiltonian systems.

\noindent{\bf Ensembles without any constraints.}---In the absence of any structure, the distribution of EE of pure random states $|\Psi\rangle$ drawn from the full Hilbert space ${\cal H}$ depends only on subsystem dimensions through the parameters $(f,L)$. In particular, the average EE, $\mu = \langle S_A \rangle$, in the asymptotic limit ($L\gg 1$) is given by
\be
\mu(L,f) \approx Lf\log d-\frac{1}{2}\delta_{f,1/2},
\label{eq:page}
\ee
which was first conjectured by Page~\cite{averagepage1993} and later proven analytically by others~\cite{randomvivo2016,weiproof2017,TypicalBianchi2019}.
The first term in the RHS of Eq.(\ref{eq:page}) is the volume-law term which scales with subsytem size $L_A = fL$, whereas the second term gives rise to the `half-qubit' shift correction for half-subsystems. {Higher statistical moments can be computed using a variety of methods\cite{randomvivo2016,weiproof2017,TypicalBianchi2019,phasetrans2010Nadal,nadal2011statistical}.} In particular, the variance of EE for pure random states, $ \sigma^{2}_{P} \approx d^{-L(1+|1-2f|)}$, is exponentially small in subsystem size, implying that the EE is typical and a single pure random state will have the Page entropy.

\noindent{\bf Ensembles constrained by one scalar charge.}---
For systems with a local scalar charge and a local Hilbert space dimension of $d=2$, it is convenient to think of $0 \leq M \leq L$ as an integer particle number, and each site only able to accommodate a maximum of one particle. The Hilbert space ${\mathcal{H}(M)}$ of states with fixed charge $M$ decomposes as a direct sum of tensor products, ${\cal H}(M) = \bigoplus_{M_A } {\cal H}_A(M_A)\otimes{\cal H}_B(M-M_A)$, where $M_A$ is within the range ${\rm max}(0, M-L_B) \le M_A \le {\rm min}(M, L_A)$. The Hilbert space dimension of ${\cal H}_A(M_A)$ is $d_{A,M_A} = \binom{L_A}{M_A}$, the Hilbert space dimension of ${\cal H}_{\rm B}(M-M_A)$ is $d_{B,M-M_A}= \binom{L-L_A}{M-M_A}$, and the total Hilbert space dimension is $\sum_{M_A} d_{A,M_A} d_{B,M-M_A} =d_M =\binom{L}{M}$. A random state with fixed total charge $|\Psi_M\rangle \in {\cal H}(M)$ can be expressed as a superposition of orthonormal basis states, $|\Psi_M\rangle = \sum_{M_A}\psi_{\alpha,\beta}^{(M_A)}|M_A,\alpha\rangle\otimes|M-M_A,\beta\rangle$, with $\psi^{(M_A)}_{\alpha,\beta}$ uncorrelated random numbers up to normalization. The index $\alpha$~($\beta$) labels the basis states in subsystem $A$~($B$) with a total charge $M_A$~($M-M_A$).

The reduced density matrix of subsystem $A$ is block diagonal, $\rho_{A|M}= \sum_{M_A} p_{M_A} \rho_{A|M_A}$, and the factors $p_{M_A}\ge 0$ are the (classical) probability distribution of finding $M_A$ particles in $A$. The entanglement entropy can be written as 
\be
S(\rho_{A|M}) = \sum_{M_A} p_{M_A} S(\rho_{A|M_A})- p_{M_A} \log p_{M_A},
\label{eq:EE_sector}
\ee
where the second term on the RHS is the Shannon entropy of the number distribution $p_{M_A}$, which captures particle number correlations between the two subsystems, and the first term captures quantum correlations between configurations with a fixed particle number.

The first few moments of the EE distribution produced by the ensemble $|\Psi\rangle \in {\cal H}(M)$ was first computed by Bianchi and Dona~\cite{TypicalBianchi2019}, see details in the Supplement. In particular, the mean entanglement entropy for `mid-spectrum' states ($M/L=1/2$) in the asymptotic limit\cite{volumebianchi2022} is given by 
\be
\mu_{M}(L,f) = Lf\log 2+\frac{f+\log(1-f)}{2}-\frac{1}{2}\delta_{f, 1/2}.
\label{eq:meanu1}
\ee
Interestingly, in addition to the volume-law term and the half-qubit shift, 
Eq.(\ref{eq:meanu1}) also exhibits a finite shift in the mean EE entropy\cite{entanglementvidmar2017} relative to the Page result~\cite{averagepage1993}.
The variance of EE scales exponentially with system size, $\sigma_{M}^2 \sim \sqrt{L}/d^{L}$, thus a typical pure random state in ${\cal H}(M)$ will have the EE in Eq.(\ref{eq:meanu1}). We emphasize that the difference between the typical EE of random states in ${\cal H}(M)$ and ${\cal H}$ is large on the exponentially small scale set by $\sigma_{M}$, see dotted and dashed-dotted lines in Fig.\ref{fig:Fig1}(a). We also note that if random vectors are real valued (i.e., GOE distributed), the mean EE is not affected at the level of $O(1)$ corrections, but the standard deviation increases by a factor of $\sqrt{2}$~\cite{kumar2011entanglement,randomvivo2016}. This is true both for constrained and unconstrained ensembles~(see Supplement). 
 
\noindent{\bf Ensembles constrained by two commuting scalar charges.}---We now consider quantum state ensembles constrained by an additional U(1) charge, which is more descriptive of typical quantum states in systems that conserve energy and magnetization. Unlike the U(1) case, there is a technical difficulty in constructing random state ensembles that have two scalar constraints if the local Hilbert space dimension is $d=2$: the symmetry operators for both charges cannot be simultaneously expressed as a sum of single site terms, thus defining a suitable basis to compute the asymptotic behavior of the EE is challenging. To address this challenge, we use a `coarse-graining' procedure in which we increase the local Hilbert space dimension as $d\rightarrow d^\lambda$, and decrease the system size as $L \rightarrow L/\lambda$, while keeping the total Hilbert space dimension constant. By `enlarging' the local Hilbert space dimension, we can express the symmetry operators for {\it both} U(1) charges in terms of local operators only. Similarly to the energy-conserving-only case, we argue that the entanglement patterns computed for large enough subsystems are insensitive to the coarse-graining procedure. 

For systems constrained by two scalar U(1) charges and a local Hilbert space dimension of $d=2$, we use $\lambda = 2$ and define the random states ensemble,
\be
|\Psi_{M,N}\rangle = \sum_{M_A,N_A}\psi_{\alpha,\beta}^{(M_A,N_A)}|M_A,N_A;\alpha\rangle\otimes \\ |M_B,N_B;\beta\rangle, 
\ee
with local Hilbert space dimension $d'=4$, $M_B=M-M_A$, and $N_B = N-N_A$. The numbers $N$ and $M$ represent the quantum numbers of each scalar charge, and each site can accommodate one particle of each flavor, see Fig.\ref{fig:Fig1}(b). 
The index $\alpha$ ($\beta$) labels basis states with $M_A$ and $N_A$ ($M_B$ and $N_B$) particles in $A$~($B$). In this case, the EE of $\rho_A$ can be expressed as 
\bea
S(\rho_{A|M,N}) = \sum_{M_A,N_A} p_{M_A,N_A} S(\rho_{A|M_A,N_A}) \\- p_{M_A,N_A} \log p_{M_A,N_A},
\eea
where each term has the same physical meaning as those in Eq.(\ref{eq:EE_sector}). When states are drawn randomly from ${\cal H}(M,N)$, the resulting distribution is computed using the same approach as in the U(1) case, and the details are discussed in the Supplement. We find that the mean EE of mid-spectrum states ($M/L=1/2$ and $N/L = 1/2$) in the asymptotic limit is given by,
\be
\mu_{M,N}(L,f) = Lf\log 2+ f+\log(1-f) -\frac{1}{2}\delta_{f, 1/2}.
\label{eq:stateu1}
\ee
In particular, the mean EE is shifted twice the value found in Eq.(\ref{eq:meanu1}) relative to the Page mean. Note that the mean EE does not change if we rescale $L \rightarrow L/\lambda$ and $d\rightarrow d^\lambda$, so long as $f$ remains constant. The variance of EE is exponentially small in systems size, $\sigma_{M,N}^2\sim L/d^L$ (see Supplement). We note that the ratio $\sigma_{M,N}/\sigma_{M}$ is $\sim L^{1/4}$; this leads to $\sigma_{M,N}$ being twice as large as $\sigma_M$ when $L=16$, as shown in Fig.\ref{fig:Fig1}(a). 
Having described the three reference ensembles, we now compare these with the eigenstate ensemble of local Hamiltonians with and without U(1) symmetry. 

\noindent{\bf Eigenstate ensembles.}---We consider the spin-1/2 Hamiltonian:
\begin{align}
    \label{eq:ModelHam}
    H = \sum_{\ell} J\left(X_{\ell}X_{\ell+1}+Y_{\ell}Y_{\ell+1}\right)&+\Delta Z_{\ell}Z_{\ell+1} \nonumber \\ 
    &+ J_2 Z_{\ell}Z_{\ell+1}Z_{\ell+2},
\end{align}
where $(X_{\ell}, Y_{\ell}, Z_{\ell})$ are the Pauli matrices.
The Hamiltonian (\ref{eq:ModelHam}) has two scalar charges, energy and total magnetization, $S_z=\sum_\ell Z_\ell$, both of which are commuting, $[H, S_z] = 0$. The Hamiltonian (\ref{eq:ModelHam}) also has multiple point symmetries, which we explicitly break. In particular, we use open boundary conditions to break translation symmetry, {and we include the boundary term $H_{\rm b}=h_b (Z_1-Z_L)$ to break inversion symmetry}. When $J_2=0$, the Hamiltonian (\ref{eq:ModelHam}) becomes the integrable XXZ chain. A finite value of $J_2$, instead, breaks integrability. We first consider the parameters $J=1$, $\Delta = J/2$, $J_2=0.6$, and $h_b = 0.25$ which we find to be the values in which eigenstates are most random.

For comparison, we also consider the Hamiltonian 
\be
H' = H + h_x\sum_\ell X_\ell,  
\label{eq:ModelHam2}
\ee 
with an additional transverse magnetic field which breaks both U(1) symmetry and integrability. When breaking U(1) symmetry, we use $h_x = 0.75$ and $J_2 = 0$.

We now analyze the EE distribution of mid-spectrum eigenstates at half-filling as a function of $L$ and $f$. In Fig.~\ref{fig:Fig2}(a), we show the EE distribution of mid-spectrum eigenstates as a function of $L$ for fixed $f=1/2$, both for the Hamiltonian (\ref{eq:ModelHam}) with U(1) symmetry (circles) and for the Hamiltonian (\ref{eq:ModelHam2}) without U(1) symmetry (squares). Each data point represents the average EE of eigenstates relative to the maximum entanglement entropy $L_A \log(2)$, and the bars indicate their standard deviation. The shaded blue area indicates the regions limited by $\mu_{M}\pm\sigma_M$ for the ensemble $|\Psi\rangle \in {\cal H}(M)$ using real-valued vectors, and the shaded red area indicates the region limited by $\mu_{M,N}\pm\sigma_{M,N}$ for the ensemble $|\Psi\rangle\in{\cal H}(M,N)$ also using real-valued vectors~(the boundaries are interpolated smoothly between the accessible $L$ values). We obtain the eigenstate distributions by choosing an energy window, $W = 20, 50, 100, 600$ of eigenstates for $L = 10,12,14,16$ around the peak density-of-states. The entanglement bipartition is placed in the center of the chain to prevent edge effects from the open boundary conditions. For the first and second moments of the EE distribution, we find excellent quantitative agreement between the eigenstate distribution and the corresponding constrained ensemble. This applies both for the Hamiltonian with a U(1) scalar charge and the Hamiltonian without additional structure.   

\begin{figure}[t]
    \includegraphics[width=\columnwidth]{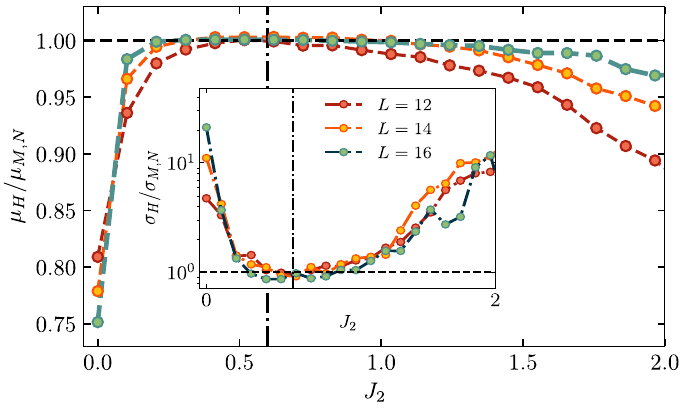}
    \caption{Finite-size scaling of the average EE of mid-spectrum eigenstates relative to the analytical prediction $\mu_{M,N}$ in Eq.(\ref{eq:stateu1}) plotted as a function of the integrability breaking parameter $J_2$. The inset shows the standard deviation of EE of mid-spectrum eigenstates relative to the analytical prediction $\sigma_{M,N}$. Excellent agreement with the constrained ensemble (at the level of the first two moments) is found in a small pocket of model parameters centered around $J_2 \approx 0.6$. Parameters used: $J=1$, $\Delta=J/2$, $h_{x}=0$, $f=1/2$.}
    \label{fig:Fig3}
\end{figure}

For subsystems with $f<1/2$, we also find that the entanglement patterns of mid-spectrum eigenstates are described remarkably well by the constrained RMT ensembles. This is shown in Fig.~\ref{fig:Fig2}(b) for fixed $L=16$, using the same number of eigenstates as in panel(a). 

As a final remark, we compare our results with previous works\cite{entanglementvidmar2017,volumebianchi2022,cheng2023typical} that have also looked at the behavior of eigenstate EE in Hamiltonian systems with an additional U(1) charge, such as that described in Eq.(\ref{eq:ModelHam}). These previous works found excellent agreement between the EE distribution of eigenstates and that of random state ensembles with a single U(1) charge {\it away} from half-filling. However, they observed deviations at half-filling that could not be explained by the same ensemble. These deviations align with our findings in Fig.~\ref{fig:Fig1}, which show large discrepancies when comparing the EE distribution of eigenstates relative to that of U(1)-constrained random states. We argue that the excellent agreement away from half-filling arises because, in this regime, a sizable $O(\sqrt{L})$ correction dominates over the $O(1)$ term~\cite{structuremurthy2019}. In contrast, at half-filling, the $O(\sqrt{L})$ term vanishes, making the $O(1)$ term the dominant contribution. In this case, it becomes important to appropriately constrain the states with two U(1) scalar charges---one for energy and the other for magnetic charge. 

\noindent{\bf Maximally chaotic systems.}---In Figs.~\ref{fig:Fig1} and \ref{fig:Fig2}, we set the next-nearest-neighbor coupling to $J_2=0.6$, as we found this value to produce the most random eigenstates. We now extend our analysis to a broader range of $J_2$ values, as shown in Fig.~\ref{fig:Fig3}. Consistent with the results in Ref.\cite{rodriguez2023quantifying} for the Mixed Field Ising Model, we observe remarkably-good agreement between the eigenstate ensemble and the (appropriately) constrained ensemble of random states specifically in small regions of parameter space near $J_2\approx 0.6$, even for relatively small system sizes ($L =12$). Outside these regions, we find finite-size deviations between the ensembles which vanish in the thermodynamic limit. These findings support the conjecture in Ref.\cite{rodriguez2023quantifying} that maximally chaotic Hamiltonians exist in small pockets of parameter space, where even for modest system sizes, the distance to the appropriately constrained ensembles is minimized. 

\noindent{\bf Discussion.}---Our work shows that {constrained} ensembles of pure random states can accurately describe increasingly finer features in the eigenstate statistics of physical systems, {with the constraints capturing the essential features of the Hamiltonian, namely spatial locality and conservation laws.} This also highlights the need to characterize new classes of random state ensembles to understand the universal properties of eigenstates in other physical systems, such as those with non-abelian symmetries~\cite{2023PRL_su2chai,2023PRB_nicolesu2,2023PRB_su2} or with integrable limits\cite{volumebianchi2022,2019JMP_gaussianstates,2021PRB_gaussianstates,2022JPA_gaussianstates}. 
On a different front, our work strengthens the claim in Ref.~\cite{rodriguez2023quantifying}, suggesting a notion of `maximal chaos' emerging in local Hamiltonians. Interesting questions to address are, what conditions are required to maximize chaos in local systems, and what are the dynamical signatures of these `maximally chaotic' Hamiltonians? 

{\bf Acknowledgements.}---We thank Vedika Khemani and Cheryne Jonay for previous collaborations and feedback on the present work, as well as Thomas Barthel, Nick Hunter-Jones, Sam Garratt, and Marcos Rigol for insightful comments. JFRN acknowledges the hospitality of the Aspen Center for Physics, which is supported by National Science Foundation grant PHY-2210452, and a grant from the Alfred P. Sloan Foundation (G-2024-22395). The numerical simulations in this work were conducted with the advanced computing resources provided by Texas A\&M High Performance Research Computing.

%

\clearpage

\renewcommand{\thefigure}{S\arabic{figure}}
\renewcommand{\theequation}{S\arabic{equation}}
\renewcommand{\thesection}{S\arabic{section}}
\setcounter{page}{1}
\setcounter{equation}{0}
\setcounter{figure}{0}
\setcounter{section}{0}

\begin{widetext}

\begin{center}
{\large\underline{\bf SUPPLEMENTARY MATERIAL} \\ {\bf Entanglement patterns of quantum chaotic Hamiltonians with a scalar U(1) charge}

\author{}}

\vspace{4mm}

Christopher M. Langlett, and Joaquin F. Rodriguez-Nieva

{\small\it $^1$Department of Physics \& Astronomy, Texas A\&M University, College Station, TX 77843}

\end{center}

\end{widetext}

\vspace{5mm}

This Supplementary Material discusses the analytical and numerical details of the results presented in the main text. Section S1 discusses the entanglement entropy~(EE) distributions generated by constrained and unconstrained random state ensembles. Section II focuses on the statistical analysis of the numerical data.

\section{Entanglement entropy distribution for ensembles of pure random states}
\label{sec:EntangleMoments}

In this section, we consider the statistical properties of the EE for different ensembles of random states, both constrained and unconstrained. We start by discussing the case of pure random states without constraints, and then discuss the cases with U(1) and U(1)$\times$U(1) constraints. In all cases, we discuss the exact analytical results as well as the asymptotic behavior, which were discussed in the main text. We finally discuss differences between ensembles drawn from complex valued distributions~(GUE) and real valued distributions~(GOE). 

\subsection{Pure random states without constraints}
\label{subsec:PageDist}

The first few moments of the EE distribution was first computed analytically in Refs.~\cite{randomvivo2016,weiproof2017}. In the following, we summarize the key results. Suppose that a system of size $L$ is partitioned into regions $A$ and $B$ of size $L_A$ and $L_B = L - L_A$, with Hilbert space dimensions $d_A = d^{L_A}$ and $d_B = d^{L_B}$, respectively. We assume without loss of generality that $L_A\le L_B$.  
A pure state $\ket{\psi}$ on the full Hilbert space $\mathcal{H}=\mathcal{H}_A \otimes \mathcal{H}_B$ can be expressed as
\begin{equation}
    \label{eq:purestate}
    \ket{\psi}=\sum_{\alpha=1}^{d_A}\sum_{\beta=1}^{d_B}\psi_{\alpha,\beta}\ket{\alpha}\otimes \ket{\beta},
\end{equation}
where $|\alpha\rangle$ ($|\beta\rangle$) are basis states of subsystem $A$~($B$), and the coefficients $\psi_{\alpha,\beta}$ represent entries of a rectangular matrix of dimension $d_A \times d_B$.
Tracing out the degrees of freedom of subsystem $B$, the reduced density matrix can be written as
\begin{align}
    \rho_A &= \sum_{\alpha,\alpha^\prime=1}^{d_A}\left[ \sum_{\beta=1}^{d_B} \psi_{\alpha,\beta}\psi^{*}_{\alpha^\prime,\beta}\right]\ket{\alpha}\bra{\alpha^\prime}.
\end{align}
The EE of $\rho_A$, $S(\rho_A)= -{\rm Tr}[\rho_A\log\rho_A]$ is obtained from the eigenspectrum $\{\lambda_\alpha\}$ of $\rho_A$, specifically, $S_A=-\sum_\alpha \lambda_\alpha \log(\lambda_\alpha)$.

Let us now consider coefficients $\psi_{\alpha,\beta}$ which are independently and identically distributed 
complex~(GUE) Gaussian variables drawn from the distribution $\mathcal{P}(\psi)= \exp\{-\text{Tr}[\psi^{\dag}\psi] \}$. This generates a distribution of EE ${\cal P}(S_A)$ with average
$\mu = \langle S_A \rangle$ given by~\cite{randomvivo2016,weiproof2017}
\begin{align}
    \label{eq:PageAverage}
    \mu = \Psi(d_A d_B +1) -\Psi(d_B +1) - \frac{d_A -1}{2d_B},
\end{align}
where $\Psi(x)=\Gamma(x)^{\prime}/\Gamma(x)$ is the digamma function, defined as the logarithmic derivative of the Gamma function.
When $L_A>L_B$, one needs to swap $d_A \leftrightarrow d_B$.
The second moment of the EE distribution, 
$\sigma=\langle S_A ^{2}\rangle -\langle S_A\rangle^{2}$, is given by:
\begin{align}
    \label{eq:ExactPageVar}
    \sigma^{2} &= \frac{d_A + d_B}{d_A d_B +1} \Psi^{\prime}(d_B + 1) - \Psi^{\prime}(d_A d_B +1)\nonumber \\
    &-\frac{(d_A -1)(d_A+2d_B-1)}{4d_B^{2}(d_A d_B +1)},
\end{align}
where $\Psi^{\prime}(x)=d\Psi(x)/dx = d^{2}[\log(\Gamma(x))]/dx^{2}$ is the first derivative of the digamma function ({\it i.e.}, the first polygamma function). 

\subsubsection{Asymptotic limit: unconstrained states}

We now consider $\mu$ and $\sigma$ in the thermodynamic limit $L\gg 1$. Assuming $f = L_A/L$ to be a finite fraction of the system, and $d_A,d_B \gg 1$, we can approximate $\Gamma(x\gg 1) \approx \log x$ and $\Gamma'(x) \approx 1/x$. In addition, the second term in Eq.(\ref{eq:PageAverage}), $(d_A-1)/2d_B$ is finite only if $L_A=L_B$. In this case, 
the first moment \eqref{eq:PageAverage} takes the form,
\be
\mu \approx Lf\log d-\frac{1}{2}\delta_{f,1/2}.
\label{seq:page}
\ee
The first term in Eq.(\ref{seq:page}) is the volume law term proportional to $L_A$, and the second term is the so-called Page correction, which is exponentially small for $f<1/2$.
The variance of the distribution ${\cal P}(S_A)$ scales as
\be
\sigma^{2} \approx \frac{1}{d_B^2} = d^{-L(1+|1-2f|)},
\label{eq:ApproxPageVar}
\ee
which is exponentially small in subsystem size. 
This implies that the EE is typical and a single pure random state will have the Page entropy.

\subsection{Pure random states with a scalar U(1) constraint}
\label{subsec:BianchiDist}

The statistical properties of EE for pure random states constrained to a U(1) symmetry sector was first derived by Bianchi and Dona~\cite{TypicalBianchi2019}. An excellent review is presented in Ref.~\cite{volumebianchi2022}. Here we quote the main results which are relevant to our work. 

Let us consider a chain of $L$ sites and a total number of $M$ particles, with $0 \le M \le L$, and each site able to accommodate a maximum of one particle. When the system is partitioned into two subsystems of sizes $L_A$ and $L_B$, the Hilbert space factors out as
\begin{equation}
    \mathcal{H}(M) = \bigoplus_{M_A={\rm max}(0,M-L_B)}^{{\rm min}(M,L_A)} \mathcal{H}_{A}(M_A) \otimes \mathcal{H}_{B}(M-M_A).
\end{equation}
The total Hilbert space dimension of each $M$ particle sector is $d_M = \binom{L}{M}$, and the Hilbert space dimensions of each subsystem
are $d_{M_A} = \binom{L_A}{M_A}$ and $d_{M_B}  = \binom{L-L_A}{M - M_A}$.

We now consider random states $|\Psi_M\rangle \in {\cal H}(M)$,
\be
|\Psi_M\rangle = \sum_{M_A}\sum_{\alpha,\beta}\psi_{\alpha,\beta}^{(M_A)}|M_A,\alpha\rangle\otimes|M-M_A,\beta\rangle,
\ee
with coefficients $\psi_{\alpha,\beta}^{(M_A)}$ which are independently and identically distributed complex~(GUE) Gaussian variables.
The reduced density matrix of subsystem $A$ is block diagonal, $\rho_{A|M}= \sum_{M_A} p_{M_A} \rho_{A|M_A}$, and the factors $p_{M_A}\ge 0$ are the (classical) probabilities of finding $M_A$ particles in $A$, $p_{M_A} = \frac{d_{M_A}d_{M_B}}{d_M}$.

The EE can be written as 
\be
S(\rho_{A|M}) = \sum_{M_A} p_{M_A} S(\rho_{A|M_A})- p_{M_A} \log p_{M_A},
\ee
where the second term on the RHS is the Shannon entropy of the number distribution $p_{M_A}$, which captures particle number correlations between the two halves, while the first term is the Page entropy for the block with $M_A$ particles in subsystem $A$. 

The first moment of EE, $\mu_M = \langle S_A \rangle_M$,  is given by\cite{TypicalBianchi2019}
\begin{align}
\label{eq:BDAverageExact}
    \mu_M =& \sum_{M_A } p_{M_A} \phi_{M_A}, \\ 
    \phi_{M_A} =& \Psi(d_M +1) - \Psi(\max(d_{M_A},d_{M_B})+1)\nonumber \\ &-\min\left(\frac{d_{M_A}-1}{2d_{M_B}},\frac{d_{M_B}-1}{2d_{M_A}}\right).
    \label{seq:phifunction}
\end{align}
In other words, the mean EE of states in a given U(1) symmetry sector is the Page entropy for all random blocks $\rho_{A|M_A}$ averaged with $p_{M_A}$.

The variance of the entanglement distribution for complex random states restricted to a symmetry sector $M$ is given by:
\begin{align}
\label{eq:BDsdevexact}
\sigma_M^2= \frac{1}{d_M +1}\bigg[\sum_{M_A}p_{M_A}\left(\phi_{M_A}^{2}+\chi_{M_A} \right) - \langle S_A \rangle_{M}^{2}\bigg],
\end{align}
where $p_{M_A}$, and $\phi_M$ are defined in the previous equations, and $\chi_M$ for $d_A \leq d_B$ takes the form,
\begin{widetext}
\be
\chi_{M_A} = 
\begin{cases}
     (d_{M_A}+d_{M_B})\Psi^{\prime}(d_{M_B}+1)-(d_M+1)\Psi^{\prime}(d_M+1) - \frac{(d_{M_A}-1)(d_{M_A}+2d_{M_B}-1)}{4d^{2}_{M_B}},\quad\text{if $d_{M_A}\le d_{M_B}$}\\
    (d_{M_A}+d_{M_B})\Psi^{\prime}(d_{M_A}+1)-(d_M+1)\Psi^{\prime}(d_M+1) - \frac{(d_{M_B}-1)(d_{M_B}+2d_{M_A}-1)}{4d^{2}_{M_A}},\quad\text{if $d_{M_A}>d_{M_B}$}.
\end{cases}
\ee
\end{widetext}

\subsubsection{Asymptotic limit: U(1) constraint}
We now consider the thermodynamic limit $L\gg 1$, and define the particle density $m = M/L$, the subsystem density $m_A = M_A/L$, and the subsystem fraction $ f = L_A/L$. In this case, the average value of $m_A$ is $\bar{m}_A = f m$. Assuming $d_A,d_B \gg 1$, we can replace the Gamma functions by $\Gamma(x\gg 1) \approx \log x$, and use the Stirling's approximation to approximate the binomial coefficients, 
\begin{align}
\log\binom{L}{M} &\approx -L[m\log(m)+(1-m)\log(1-m)]\nonumber \\
&+ \frac{1}{2}\log\left[\frac{L}{M(L-M)}\right] + \frac{1}{2}\log(2\pi).
\label{eq:stirling}
\end{align}

Using these approximations, the terms $p_{M_A}$ and $\phi_{M_A}$ in Eq.(\ref{eq:BDAverageExact}) can be expressed in powers of $(m_A-\bar{m}_A)$ around $\bar{m}_A$ in order to compute $\mu_M$ up to O(1) terms.
The probability distribution of $m_A$ is Gaussian-distributed, 
\begin{align}
\label{eq:guassian}
    &p_{m_A} = \nonumber \\
    &\sqrt{\frac{2\pi L}{m(1-m)f(1-f)}} \exp \bigg[-\frac{L}{2}\frac{(m_A-fm)^2}{m(1-m)f(1-f)} \bigg],
\end{align}
peaked at $\bar{m}_A=f{m}$,
and has a standard deviation given by $\sigma_{m_A} = \sqrt{m(1-m)f(1-f)/L}$.

Using $\Gamma(x\gg 1)\approx \log x$, the term $\phi_{M_A}$ in Eq.(\ref{seq:phifunction}) can first be approximated as 
\be
\phi_{M_A} = \log\left(\frac{d_{M}}{d_{M_B}}\right) - \frac{1}{2}\delta_{f,1/2}\delta_{m,1/2}.
\ee
Using Stirling's approximation, $\phi_{M_A}$ is found to be
 \begin{widetext}
\begin{align}
\phi_{M_A} \approx -L[m\log(m)+(1-m)\log(1-m)]&+ L[m_B\log(m_B) +(1-f-m_B)\log(1-f-m_B)-(1-f)\log(1-f)]\nonumber \\ &+\frac{1}{2} \log \bigg[\frac{m_B(1-f-m_B)}{m(1-m)(1-f)}\bigg]-\frac{1}{2}\delta_{f,1/2}\delta_{m,1/2}.
\end{align}
\end{widetext}

The last step is integrating $\phi_{m_A}$ using the Gaussian distribution $p_{m_A}$. In this work,  we primarily focus on mid-spectrum states, thus, we set $m=1/2$.
In this case, the mean EE up to O(1) terms is given by 
\be
\mu_{M}(L,f) = Lf\log(d)+\frac{f+\log(1-f)}{2}-\frac{1}{2}\delta_{f, 1/2}.
\label{eq:BDaverageasymptotic}
\ee

The same procedure can be applied to the standard deviation $\sigma_M$ in Eq.(\ref{eq:BDsdevexact}). We note that the term in brackets in Eq.(\ref{eq:BDsdevexact}) is O(1), thus the system size dependence of $\sigma_M^2$ is dominated by the term $d_M$ in the denominator. The value of 
$\sigma_M^2$ can then be approximated as
\be
\sigma_{M}^2 \approx \frac{1}{d_M}  \sim \frac{\sqrt{L}}{2^{L}},
\label{eq:BDstdapprox}
\ee
where the numerical prefactor of $\sigma_M^2$ depends on $f$. Unlike that Page variance $\sigma^2$ in Eq.(\ref{eq:ApproxPageVar}), however, here we note that $\sigma_M^2$ is a factor $\sqrt{L}$ larger than $\sigma^2$. This is because $d_M = { L \choose M }$ is a factor $\sqrt{L}$ smaller than $d^L$ in the asymptotic limit. 

\subsection{Pure random states with two scalar U(1)$\times$U(1) constraints.}

We now discuss the distribution of EE of pure random states with a U(1)$\times$ U(1) constraint, defined by a Hilbert space, $\mathcal{H}(M, N)$ with a fixed number of particles $M$ and $N$.
Let us consider a system of $L$ sites with a local Hilbert space dimension $d=4$, with the local basis states spanned by either an empty site, a site occupied by particle-$1(2)$ or occupied by both particles~[see Fig.~\eqref{fig:Fig1}(b) in the main text].
The Hilbert space dimension of a symmetry sector with $(M, N)$ particles is 
\begin{equation}
    \label{eq:Dim}
    d_{M,N}=\binom{L}{M}\binom{L}{N},
\end{equation}
with $\sum_{M, N =0}^{L}\binom{L}{M}\binom{L}{N}=4^{L}$ by Vandermonde's identity. 
We next bipartition the system into $A$ and $B$ subregions of size $L_A$ and $L_B = L-L_A$.
In this case, the Hilbert space of the $(M,N)$ symmetry sector decomposed as
\begin{align}
    \label{eq:HilbertSpaceSymmetry}
    \mathcal{H}(M,N) =\nonumber \\ \bigoplus_{M_{A},N_{A}}&\mathcal{H}_A(M_{A}, N_{A}) \otimes \mathcal{H}_B(M-M_{A}, N-N_{A}), 
\end{align}
where the Hilbert space dimension of subsystem $A$ with $(M_A,N_A)$  particles is given by $d_{M_A,N_A}=\binom{L_A}{M_A}\binom{L_A}{N_A}$,
and the Hilbert space dimension of subsystem $B$ with $(M_B,N_B)$ particles is given by $d_{M_B,N_B}=\binom{L-L_A}{M-M_A}\binom{L-L_A}{N-N_A}$. 

Let us consider a state $\ket{\Psi} \in \mathcal{H}(M,N)$
\begin{align}
    \label{eq:PureState}
    \ket{\psi} = \sum_{M_A, N_A} \sum_{\alpha, \beta} \psi^{(M_A,N_A)}_{\alpha, \beta}&\ket{M_A, N_A,\alpha}\nonumber \\ 
    &\otimes \ket{M-M_A, N-N_A,\beta},
\end{align}
with the coefficients $\psi_{\alpha,\beta}^{(M_A,N_A)}$ independently and identically distributed complex~(GUE) Gaussian variables. Here $|M_A,N_A,\alpha\rangle$ ($|M_B,N_B,\beta\rangle$) are basis states of subsystem $A$ ($B$) with $(M_A,N_A)$ [($M_B,N_B$)] particles. The reduced density matrix of the above state over subsystem $A$ is of block diagonal form, namely,
\begin{equation}
    \rho_A =\sum_{M_A, N_A}p_{M_A,N_A}\rho_{A|M_A,N_A}.
\end{equation}
The values $p_{M_A,N_A}$ come from normalizing the reduced density matrix within in each $(M_A, N_A)$ sector and satisfy $\sum_{M_A,N_A} p_{M_A,N_A} = 1$.
The factors $p_{M_A,N_A}$ are interpreted as the classical probability of finding $M_A$ and $N_A$ particles over the subregion $A$.
Moreover, the entanglement entropy becomes
\begin{align}
    S(\rho_A) = \sum_{M_A, N_A} p_{M_A,N_A} &S(\rho_{A|M_A,N_A})\nonumber \\ &-p_{M_A,N_A}\log(p_{M_A,N_A}),
\end{align}
where the second term is the Shannon entropy which captures classical correlations between particles of the subsystems, while the first term captures quantum correlations.
Importantly, since the two $U(1)$ charges commute, we assume that $p_{M_A,N_A} = p_{M_A}p_{N_A}$ are independent probabilities for each flavor. 
In addition, each block $\rho_{A|M_A,N_A}$ with $(M_A,N_A)$ particles is assumed to be a random matrix. In this case, $S(\rho_{A|M_A,N_A})$ is the Page entropy (\ref{eq:PageAverage}) for pure random strained constrained to $(M_A,N_A)$ particles in subsystem $A$.
Thus, the average EE $\mu_{M,N} = \langle S_A\rangle_{M,N}$ for pure random states in the $(M,N)$ symmetry sector is the Page entropy for each block $\rho_{A|M_A,N_A}$ averaged with the distribution $p_{M_A,N_A}$, 
\begin{align}
    \label{eq:AvEntTwoCharge}
    \mu_{M,N} &= \sum_{M_A, N_A} \frac{d_{M_A,N_A}d_{M_B,N_B}}{d_{M,N}} \phi_{M_A,N_A}\\
    \phi_{M_A,N_A} &= \Psi(d_{M,N}+1)-\Psi(\max(d_{M_A,N_A},d_{M_B,N_B})+1)\nonumber \\
    &-\min\bigg[ \frac{d_{M_A,N_A}-1}{2d_{M_B,N_B}},\frac{d_{M_B,N_B}-1}{2d_{M_A,N_A}} \bigg]\nonumber,
\end{align}
with $\Psi(z)=\Gamma^{\prime}(z)/\Gamma(z)$ the Digamma function, and $d_{A(B)}$ defined above.
The variance of the entanglement distribution for complex random states restricted to a symmetry sector $(M,N)$ is given by:
\begin{align}
\label{eq:sdevTwoCharge}
\sigma_{M,N}^2 &= \frac{1}{d_{M,N} +1}\bigg[\sum_{M_A,N_A}p_{M_A,N_A}\left(\phi_{M_A, N_A}^{2}+\chi_{M_A,N_A} \right)\nonumber \\
&- \langle S_A \rangle_{M,N}^{2}\bigg],
\end{align}
where $p_{M_A,N_A}=\frac{d_{M_A,N_A}d_{M_B,N_B}}{d_{M,N}}$, and $\phi_{M_A,N_A}$ are defined in the previous equations, and $\chi_{M_A,N_A}$ takes the form,
\begin{widetext}
\be
\chi_{M_A,N_A} = 
\begin{cases}
     (d_{M_A,N_A}+d_{M_B,N_B})\Psi^{\prime}(d_{M_B,N_B}+1)-(d_{M,N}+1)\Psi^{\prime}(d_{M,N}+1) - \frac{(d_{M_A,N_A}-1)(d_{M_A,N_A}+2d_{M_B,N_B}-1)}{4d^{2}_{M_B,N_B}},\\
     \\
    (d_{M_A,N_A}+d_{M_B,N_B})\Psi^{\prime}(d_{M_A,N_A}+1)-(d_{M,N}+1)\Psi^{\prime}(d_{M,N}+1) - \frac{(d_{M_B,N_B}-1)(d_{M_B,N_B}+2d_{M_A,N_A}-1)}{4d^{2}_{M_A,N_A}}.
\end{cases}
\ee
\end{widetext}
Where case 1: $d_{M_A,N_A}\le d_{M_B,N_B}$, and case 2: $d_{M_A,N_A}>d_{M_B,N_B}$.

\subsubsection{Asymptotic limit: U(1)$\times$U(1) constraint}
We now consider the asymptotic limit $L\gg 1$, and define the particle densities $m = M/L$, $n = N/L$, $m_A = M_A/L$, $n_A = N_A/L$, and the subsystem fraction as $ f = L_A/L$. In this case, the average value of $m_A$ and $n_A$ is $\bar{m}_A = f m$ and $\bar{n}_A = fn$. Because $d_A,d_B \gg 1$, we can also replace the Gamma functions by $\Gamma(x\gg 1) \approx \log x$, and use the Stirling's approximation in Eq.(\ref{eq:stirling}) to approximate the binomial coefficients. For each symmetry sector, the probability distributions acquire a Gaussian form like in Eq.(\ref{eq:guassian}). In addition, the asymptotic form of the term $\phi_{M_A,N_A}$ 
is 
\be
\phi_{M_A,N_A} = \log\left(\frac{d_M}{d_{M_B}}\frac{d_N}{d_{N_B}}\right)-\frac{d_{M_A}d_{N_A}-1}{2d_{M}d_{N}}.
\ee
Putting everything together into the expression for $\mu_{M,N}$, we obtain
\bea
\mu_{M,N} = \sum_{M_A,N_A}p_{M_A}p_{N_A}\left[\log\left(\frac{d_M}{d_{M_B}}\frac{d_N}{d_{N_B}}\right)-\right. \\ 
\left.\frac{d_{M_A}d_{N_A}-1}{2d_{M}d_{N}}\right]
\eea
Because we assume the distributions $p_{M_A}$ and $p_{N_A}$ to be independent, the average entanglement entropy separates into two equivalent expressions
\bea
\langle S_A \rangle_{M,N} = \sum_{M_A}p_{M_A}\log\left(\frac{d_M}{d_{M_B}}\right) + \sum_{N_A}p_{N_A}\log\left(\frac{d_N}{d_{N_B}}\right) \\ - \frac{1}{2}\delta_{f,1/2}\delta_{m,1/2}\delta_{n,1/2},
\eea
after using $\sum_{M_A}p_{M_A} = 1$, and $\sum_{M_B}p_{M_B}=1 $. Each of the sums can be solved in exactly the same way as described in the previous section. 

When evaluating for pure random states with $m = n = 1/2$, the average EE in the asymptotic limit is 
\begin{align}
\label{eq:FirstOrder}
    \mu_{M,N} \approx & Lf\log(d) + f +\log(1-f)-\frac{1}{2}\delta_{f,1/2}.
\end{align}
In this case, we find twice the EE shift (relative to the Page value) compared to that found in Eq.(\ref{eq:BDaverageasymptotic}).

\begin{figure*}[t]
    \includegraphics[width=\textwidth]{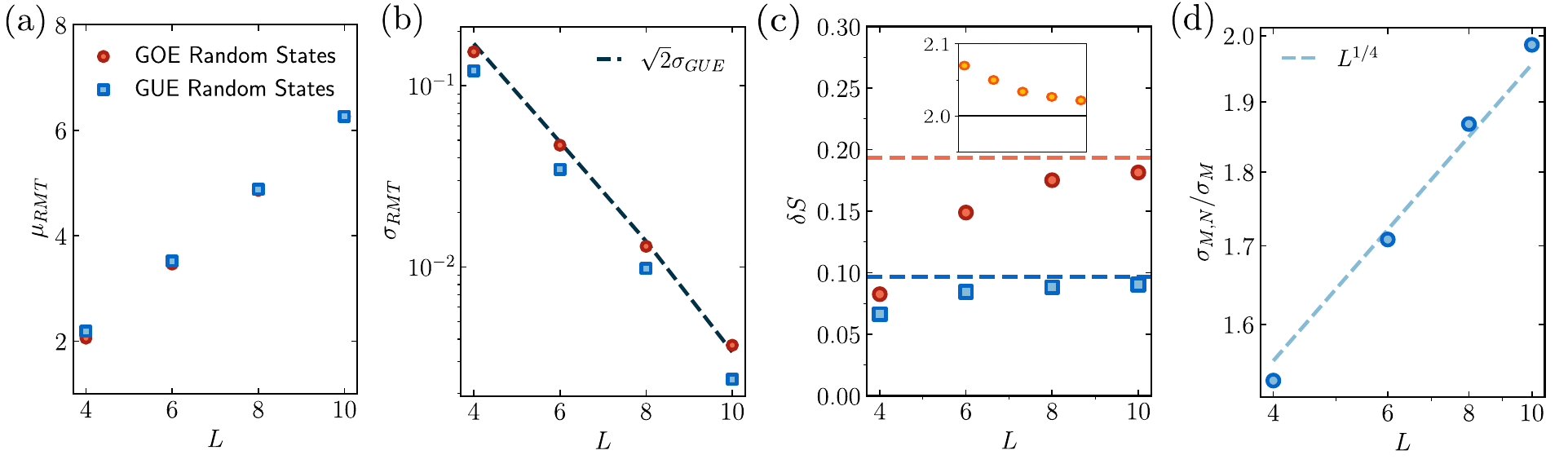}
    \caption{
    (a) System-size dependence of the average entanglement entropy of pure random states drawn from a GUE~(blue) and GOE~(red) distribution in the presence of a U(1)$\times$U(1) constraint. In the asymptotic limit, both are converging to the same result. 
    (b) Numerical data for the standard deviations of the entanglement entropy using the same random pure states generated in panel (a). The dotted line shows that the standard deviation of EE of real random states is a factor $\sqrt{2}$ larger than that of complex random states~\cite{rodriguez2023quantifying}. 
    (c) The deviation from maximal EE, $S_{max}=L\log(2)-0.5$ at $f=1/2$ for both U(1)$\times$U(1)~(red) and U(1)~(blue) constraints. The U(1)$\times$U(1) case converges to $|f+\log(1-f)|=0.19$, which is two times larger than $|(f+\log(1-f))/2|=0.096$ for the U(1) constraint. 
    Inset: The ratio of the red and blue points showing convergence to the value 2.
    (d) Standard deviation of EE for pure random states with a U(1)$\times$U(1) constraint relative to that of pure random states with a U(1) constraint. Parameters used: The number of random states for $L=4,6,8,10$ are  $10^{6},10^{6}, 10^{5}, 10^{4}$.
    }
    \label{fig:SuppFig1}
\end{figure*}

In the main text, we argue that, in the case of systems with two local scalar charges and Hilbert space dimension $d = 2$, one can `coarse-grain' the system as $L' = L/\lambda$ and $d' = d^\lambda$ and, effectively, one obtains the same entanglement patterns as in the unrescaled system. We note that Eq.(\ref{eq:FirstOrder}) is invariant to rescaling. In particular, the volume law term remains invariant upon rescaling, $(L/\lambda)f\log(d^\lambda) = Lf\log(d)$, 
and the same occurs for the O(1) terms so long as the subsystem fraction $f$ remains constant. 
Finally, we comment on the standard deviation of the EE in a system with a U(1)$\times$U(1) charge in the case of half-filling. Similarly to the U(1) case, the variance scales with the inverse of the Hilbert space size of the symmetry sector $(M,N)$, thus
\be
\sigma_{M,N}^2 \approx \frac{1}{d_{M,N}} \sim \frac{L}{d^L},
\ee
Unlike the variance $\sigma_M^2$ in Eq.(\ref{eq:BDstdapprox}), however, here we note that $\sigma_{M,N}^2$ is a factor ${L}$ larger than $d^L$ due to the midspectrum symmetry sector $d_{M,N}$ being a factor $1/L$ smaller than the dimension of the full Hilbert space. As such, we find that that $\sigma_{M,N}/\sigma_M \sim L^{1/4}$, see Fig.\ref{fig:SuppFig1}(d).

\section{Numerical Methods.}
This section gives details on the numerical procedures used to analyze the numerical data in the main text. First, we go over constructing random pure states drawn from either a GOE or GUE ensemble and discuss how the second moment changes depending on which random matrix ensemble is used. Second, we discuss how to pick a proper window size of infinite temperature states before finite-temperature state become relevant.

\subsection{Random Pure State Entanglement}

In Fig.~\eqref{fig:SuppFig1}, we numerically compare the first two moments of the entanglement distribution for different system sizes using the random states in Eq.~\eqref{eq:PureState} where the $\psi_{\alpha,\beta}^{(M_A,N_A)}$ are drawn either from a GUE or GOE ensemble.
While exact expressions for the first moments differ between GUE and GOE, they are asymptotically the same in the thermodynamic limit~\cite{randomvivo2016,vivo2010entangled,kumar2011entanglement}.
Fig.~\ref{fig:SuppFig1}(a) confirms 
that there is little difference between the average entanglement values.
This is consistent with the results found in Ref.\cite{2023PRB_su2}. 
Figure~\ref{fig:SuppFig1}(b) also shows that the standard deviation of $\sigma_{M,N}$(GUE) is a factor $\sqrt{2}$ larger than $\sigma_{M,N}$(GOE) (shown by the blue dashed line). 
The additional prefactor is due to averaging over both the real and complex parts of the GUE ensemble.

In Fig.~\ref{fig:SuppFig1}~(c) we isolate the $O(1)$ correction to the average EE for both a single and two scalar charges using pure random states.
Specifically, we compute EE relative to the Page entanglement entropy, $S_A\log(2)-0.5$.
We find that, as the system size grows in size, the difference approaches either $|f+\log(1-f)|$ for two scalar charges~(red) or $|f+\log(1-f)|/2$ for a single charge~(blue).
The variance of the EE 
distribution are exponentially small~[see above], but the standard deviation has a polynomial scaling with system size relative to $\sigma$, see Eq.(\ref{eq:ExactPageVar}), which we illustrate in Fig.~\ref{fig:SuppFig1}(d).

\begin{figure*}[t]
    \includegraphics[width=\textwidth]{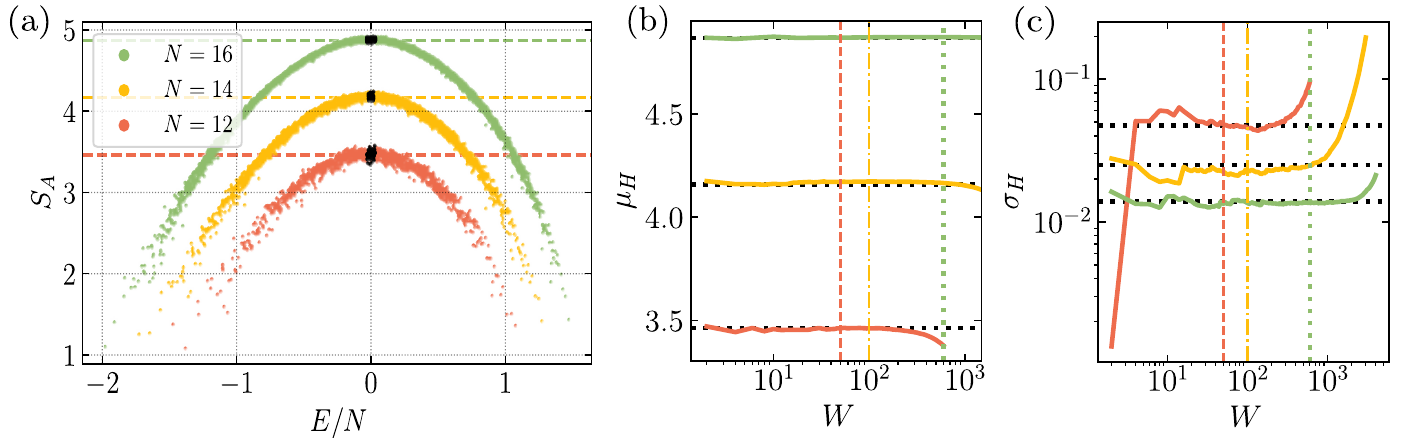}
    \caption{
    (a) Eigenstate entanglement entropy of the Hamiltonian with two U(1) charges. The black points are the eigenstates used in averaging for that system size. The dashed horizontal lines are the average values from random pure states.
    (b) Mean and (c) standard deviation of the entanglement distribution as a function of the window size, $W$. Shown with vertical lines is the number $W=50,100,600$ used in averaging in the main-text. Black horizontal lines are the average entanglement from pure random state averages.
    Parameters in all panels: $J=1$, $\Delta=J/2$, $J_{2}=0.62$, $J_{x}=0$ for $f=1/2$.
    }
    \label{fig:SuppFigA2}
\end{figure*}

\subsection{Window Size Dependence on Distribution Moments}
\label{sebsec:WindowAnalysis}

For the Hamiltonian in Eq.(\ref{eq:ModelHam}) of the main text, when computing the first two moments of the EE distribution for mid-spectrum eigenstates, it is necessary to take a finite energy window $\Delta E$ in which to take samples of $S_A$, see Fig.~\ref{fig:SuppFigA2}(a). 
If the energy window is too small, then a statistically small number of states will be available for sampling, thus leading to large error bars.
On the other hand, if the window is too large, 
then finite-energy eigenstates with low entanglement will skew the distribution and increase its variance, see Figs.\ref{fig:SuppFigA2}(c). 
Because of typicality, we argue only a few eigenstates are necessary to quantify the mean and standard deviation.
In Fig.~\ref{fig:SuppFigA2}~(a) we plot the eigenstate entanglement of the full spectrum for different system sizes, highlighting the set of eigenstates used in the main text to construct the microcanonical average [indicated with vertical lines in (b) and (c)].

\end{document}